\documentclass[pra,twocolumn]{revtex4}

\pdfoutput=1

\usepackage{graphicx}

\begin{document}

\title{Anisotropy in the Interaction of Ultracold Dysprosium } 

\author{Svetlana Kotochigova}
\email[Corresponding author: ] {skotoch@temple.edu}
\author{Alexander Petrov}
\altaffiliation{Alternative address: St.~Petersburg Nuclear Physics Institute, Gatchina, 188300; Department of Physics, St.Petersburg State University, 198904, Russia}
\affiliation{Department of Physics, Temple University, Philadelphia, PA 19122-6082, USA} 

\begin{abstract}
The nature of the interaction between ultracold atoms with a large orbital
and spin angular momentum has attracted considerable attention. It
was suggested that such interactions can lead to the realization
of exotic states of highly correlated matter. Here, we report on a
theoretical study of the competing anisotropic dispersion, magnetic
dipole-dipole, and electric quadrupole-quadrupole forces between two
dysprosium atoms. Each dysprosium atom has an orbital angular momentum $L=6$ and magnetic
moment $\mu=10\mu_B$. We show that the dispersion coefficients of the
ground state adiabatic potentials lie between 1865 a.u. and 1890 a.u.,
creating a non-negligible anisotropy with a spread of 25 a.u. and that the
electric quadrupole-quadrupole interaction is weak compared to the other
interactions.  We also find that for interatomic separations $R< 50\,a_0$
both the anisotropic dispersion and  magnetic dipole-dipole potential are
larger than the atomic Zeeman splittings for external magnetic fields
of order  10 G to 100 G. At these separations spin exchange can occur.
We finish by describing two scattering models for inelastic spin exchange.
A universal scattering theory is used to model loss due to the anisotropy
in the dispersion and a  distorted-wave-Born theory is used to model
losses from the magnetic dipole-dipole interaction for the $^{164}$Dy isotope. 
These models find loss rates that are the same order of magnitude as the experimental value.
\end{abstract}

\pacs{ }

\maketitle

\section{Introduction}
In recent years significant effort has been devoted to the characterization of the interactions with submerged-shell 
3d-transition-metal and 4f-rare-earth atoms
\cite{Hensler2003,Hancox2004,Hancox2005,Krems2005,Stuhler2005,Harris2007,Lahaye2007,Connolly2010}.
These atoms have an electronic configuration with an unfilled inner
shell shielded by a closed outer shell.  They also tend to have a
large magnetic moment due to a large number of unpaired electrons, which
presents opportunities to explore the effect of anisotropic magnetic dipole-dipole
interactions between them. Long-range dipolar interactions create conditions for
realizing novel quantum states of highly correlated ultracold atomic matter \cite{Lahaye2007,Fregoso2009}.  
This physics  complements that proposed with  ultracold polar molecules, another
system in which exotic quantum phases are predicted \cite{Wilson2009,Baranov2011,Wang2006,Lewenstein2006}.
Here,  dipole-dipole forces originate from a non-zero electric
dipole moment.  Unlike for magnetic atoms, however, the electric dipole
moment must be induced by an external electric field.

Submerged shell atoms are expected to have significantly suppressed
inelastic, energy-releasing spin-exchange collisions because of
shielding caused by the closed outer-shell electrons.  This effect was
first predicted and demonstrated for collisions between submerged-shell atoms with helium
\cite{Aleksandrov1983,Hancox2004,Hancox2005,Krems2005}. 
The suppression of inelastic loss with the He atom
indicates  that there is no collisional anisotropy.
The spherically symmetric He atom can not take up angular
momentum from the submerged-shell atom.

Recent measurements
of the spin-exchange rates between two submerged-shell atoms,
however, have seen no suppression and, in fact, the rate coefficients
are of the same order of magnitude as for non-submerged shell atoms
\cite{Harris2007,BLev2010,Connolly2010}. 
A possible explanation for this phenomena, given in
Ref.~\cite{Connolly2010}, is that most submerged-shell atoms have a
non-zero orbital electron angular momentum $L$.  This leads a non-zero
electrostatic quadrupole moment and anisotropic quadrupole-quadrupole
interaction that, in principle, can cause substantial losses.

In this paper we propose and discuss another mechanism that leads
to losses.  We will show that the large loss rate of order $10^{-10}$
cm$^3$/s, observed in \cite{Harris2007,BLev2010,Connolly2010}, might
have been due to anisotropy in the dispersion forces at short
inter-atomic separations. This anisotropy is also induced by the nonzero
$L$.  We study this new mechanism of spin-exchange for the submerged-shell atom with
the largest magnetic moment, dysprosium.  It has an unfilled 4f$^{10}$
shell lying beneath a filled 6s$^2$ shell leading to a large orbital,
$L$ = 6, and total, $j$ = 8, angular momentum.  Its ground  $^5{\rm I}_{8}$ state has a magnetic
moment of $\mu$ = 10$\mu_B$, where $\mu_B$ is the Bohr magneton.
Only recently, the first laser cooling and trapping experiments of a large
number of dysprosium atoms have been reported \cite{BLev2010}. The first
measurements of inelastic collisional rates in this study suggest that
anisotropy in the inter-atomic forces plays a significant role.

The paper is organized as follows. We first analyze the 
isotropic and anisotropic dispersion interaction between two Dy atoms in Section~\ref{secdisp} and compare it
with the magnetic dipole-dipole and electrostatic quadrupole-quadrupole interactions.
The  dispersion coefficients are calculated from atomic transition frequencies and
dipole moments. The quadrupole moment of Dy is determined from a multi-configuration electronic structure calculation.
In Section~\ref{strength} we  study  the relative strength of the interactions in the presence of an external magnetic
field and rotation.
In Section \ref{rates} we use these interactions to find the first estimates of the spin-exchange loss
rates and compare with experimental results. 

\section{Relative strength of interaction forces between ground state Dy atoms}\label{secdisp}

The theoretical calculation of the ground state Dy$_2$ potentials and their
dispersion parameters is challenging due to the complexity
of the spin structure of the ground-state $^5{\rm I}_{8}$ Dy atom. For example, there are 81 gerade
and 72 ungerade potentials that dissociate to the $^5{\rm I}_{8}$
+ $^5{\rm I}_{8}$ limit. In spite of this complexity we have begun to
calculate the van der Waals $C_6$ coefficients for two interacting Dy atoms.  

For two colliding atoms we can define the angular momentum ${\vec J}={\vec
\jmath}_{1}+{\vec \jmath}_{2}$, its projection $M$ along the direction of the external magnetic
field $\vec B$, and its projection $\Omega$ along the internuclear axis.
For this relativistic molecule the adiabatic Born-Oppenheimer
(BO) potentials are labeled by $\Omega_{\sigma}^{\pm}$, where $\sigma=g/u$ for
gerade and ungerade states, respectively. Gerade (ungerade) symmetry
corresponds to superpositions of even (odd) values of $J$. The superscript $\pm$ is
only relevant for $\Omega=0$ states.
For each $\Omega$ there are $17-|\Omega|$ adiabatic Born-Oppenheimer
(BO) potentials combined.

\subsection{Electrostatic dispersion interaction}

We describe the dispersion interaction potential for two ground-state
atoms in state $|j_1m_1, j_2m_2\rangle$  using  degenerate second-order
perturbation theory similar to that given by Ref.~\cite{Stone}. 
The magnetic quantum numbers $m_1$ and $m_2$ are projections along
the internuclear axis of the total atomic angular momenta $\vec j_1$
and $\vec j_2$ for the two atoms, respectively.  Here $j_1=j_2=8$.
(In this section we break with convention and use roman symbols for
atomic projection quantum numbers on the internuclear axis.)
Matrix elements of the dispersion potentials are
\begin{eqnarray}
\lefteqn{ \langle j_1 m_1, j_2 m_2 | U_{\rm disp} | j_1 m'_1, j_2 m'_2 \rangle =  -\frac{C_6(m_1m_2, m'_1m'_2)}{R^6}   }
\nonumber\\
& =& \sum_{n_a j_a m_a\atop n_b j_b m_b} \frac{1}{(E_1+E_2)- (E_{n_aj_a}+E_{n_bj_b}) } \\ \nonumber
& & \quad \quad\quad \langle j_1m_1, j_2m_2 | \hat{V}_{dd} | n_aj_am_a, n_bj_bm_b\rangle  \label{potential} \\ 
& & \quad \quad\quad \quad\quad \times \langle n_a j_a m_a, n_bj_bm_b | \hat{V}_{dd} | j_1m'_1, j_2m'_2 \rangle \ ,
  \nonumber 
\end{eqnarray}
where the $C_6(m_1m_2, m'_1m'_2)$ form a matrix of dispersion coefficients, $R$ is the 
separation between the atoms, the sums are over all electronic states 
$| n_aj_am_a, n_bj_bm_b\rangle$ of atoms $a$ and $b$  excluding states with energies $ E_{n_a j_a}$
and $E_{n_b j_b}$ equal to the ground state energies $E_1$ and $E_2$. The operator $\hat V_{dd}$ is 
the dipole-dipole interaction Hamiltonian \cite{Stone}
\begin{equation}
              \hat V_{dd}(\vec R) =\frac{1}{4\pi\epsilon_0} 
                                    \frac{(\vec {d_1} \cdot \vec {d_2}) - 3d_{1z}d_{2z}}{R^3} \,
\label{disp}
\end{equation}
where $\epsilon_0$ is the electric constant, ${\vec d}_1$ and $\vec d_2$ are the electric dipole operators for the two atoms, and $d_{1z}$ and $d_{2z}$ 
are their components  along the internuclear axis.

Using the Wigner-Eckart theorem we write the matrix $C_6$  as
\begin{eqnarray}
C_6 (m_1m_2, m'_1m'_2) = \sum_{j_aj_b} K_{j_aj_b}^{j_1j_2}
A^{j_1j_2j_aj_b}_{m_1m_2,m'_1m'_2},
\label{C_6}
\end{eqnarray}
where
\begin{eqnarray}
\nonumber
& &A^{j_1j_2j_aj_b}_{m_1m_2,m'_1m'_2}  =  \sum_{m_a,m_b} (1+\delta_{m_1,m_a} )(1+\delta_{m'_1,m_a} )  \\ \nonumber
\nonumber
& &
 \left(\begin{array}{ccc}
                          j_1 &    1      & j_a \\ \nonumber
                         -m_1 & (m_1-m_a) & m_a
             \end{array}\right)
\left(\begin{array}{ccc}
                          j_2 &    1      & j_b \\
                         -m_2 & (m_2-m_b) & m_b
             \end{array}\right) \\
&&
\left(\begin{array}{ccc}
                          j_a &    1      & j_1 \\
                        -m_a & (m_a-m'_1) & m'_1
             \end{array}\right) 
\left(\begin{array}{ccc}
                          j_b &    1      & j_2 \\ \nonumber
                         -m_b & (m_b-m'_2) & m'_2
             \end{array}\right)\,,
\end{eqnarray}
and
\begin{eqnarray}
K_{j_aj_b}^{j_1j_2} =\left(\frac{1}{4\pi\epsilon_0}\right)^2  \sum_{n_a,n_b}
\frac{\left|\left<j_1||d_1||n_aj_a\right>\left<j_2||d_2||n_bj_b\right>\right|^2}
{(E_{n_aj_a}+E_{n_bj_b}) - (E_1+E_2)}\,. \nonumber
\end{eqnarray}
Note that the $A^{j_1j_2j_aj_b}_{m_1m_2,m'_1m'_2}$ conserve the molecular projection
$\Omega=m_1+m_2=m'_1+m'_2$ and are
independent on atomic transition frequencies and dipole moments. For this homonuclear molecule gerade/ungerade
symmetry states are most conveniently constructed by transforming to states of total $\vec J$. That is to states
$|(j_1j_2)J\Omega\rangle$ and noting that even(odd) $J$ states have gerade(ungerade) symmetry.

There are six independent $K_{j_aj_b}^{j_1j_2} $ for
two Dy $^5$I$_8$ atoms as the selection rules of the electric dipole operator requires that
$|j_1-1|\le j_a \le j_1+1$ and $|j_2-1|\le j_b \le j_2+1$. For homonuclear dimers the $K_{j_aj_b}^{j_1j_2} $
is symmetric under interchange of $j_a$ and $j_b$.
We have determined $K_{j_aj_b}^{j_1j_2} $ using 62 experimental transition frequencies and oscillator strengths 
from the ground to various excited states of the Dy atom
\cite{Exp_f}. Table \ref{Ks} lists the values of $K_{j_aj_b}^{j_1j_2} $.

\begin{figure} \includegraphics[scale=0.3]{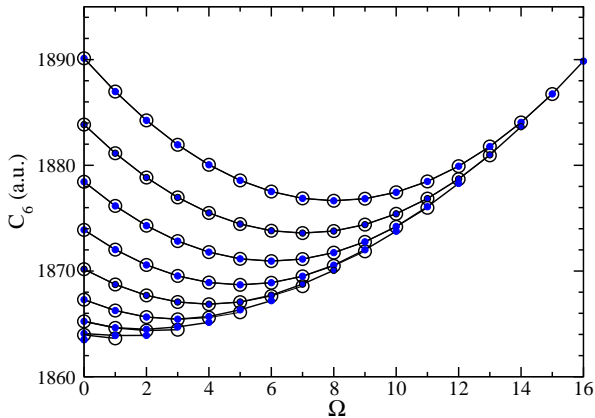} 
\caption{Gerade (filled circles) and ungerade (open circles) adiabatic $C_6$ coefficients for 
the interaction between two ground $^5{\rm I}_{j=8}$ state Dy atoms as a function of 
the projection $\Omega$ of the total angular momentum $\vec J$ on the interatomic axis.
The difference between the dispersion coefficients for the gerade/ungerade symmetry is 
small and invisible on the graph. A larger C$_6$ coefficient implies a deeper 
Born-Oppenheimer potential.} 
\label{C6} 
\end{figure}

The adiabatic dispersion potentials and, thus, the long-range of the Born-Oppenheimer potentials
are found by the diagonalizing
the $C_6$ matrices for each $\Omega^{\pm}_{g/u}$.  Figure~\ref{C6} shows the adiabatic gerade and ungerade 
$C_6$ coefficients as a function of the projection quantum number $\Omega$ of the total
angular momentum $J$ on the interatomic axis. The number of of adiabatic $C_6$
values is smaller for larger $\Omega$. In fact, for $\Omega$ = 16 there is only
one potential. It has gerade symmetry. In total there are 81/72
dispersion coefficients corresponding to the ground state gerade/ungerade
potentials. The coefficients in Fig.~\ref{C6} show a smooth nearly parabolic behavior
with the projection number $\Omega$. This $\Omega$ dependence is a consequence of the
anisotropic coupling of the open $f$-shell electrons of the two atoms. As a result the
interaction energy depends on the relative orientation of the atoms.

\begin{table}[b]
\caption
{The  $K_{j_aj_b}^{j_1j_2} $ matrix elements in atomic units for the dipole transitions 
from $j_1= j_2 = 8$ to $j_a$, $j_b$ = 7, 8, or 9 for  two interacting Dy atoms.
}
\begin{tabular}{l|c|c|c}
\hline
$j_a$/$j_b$&  7     &         8       &      9   \\
\hline
7    &   71528.597  &      81313.663  &      88173.833\\
8    &   81313.662  &      92438.922  &      100240.311\\
9    &   88173.833  &      100240.311  &      108705.654\\
\end{tabular}
\label{Ks}
\end{table}

\subsection{Magnetic dipole-dipole and electrostatic quadrupole-quadrupole interaction} 

The matrix elements of the magnetic dipole-dipole interaction between two magnetic dipoles $\vec \mu=g_j\mu_B\vec j $  is 
\begin{eqnarray}
    \lefteqn{ \langle j_1 m_1, j_2 m_2 | U_{\rm mdd}| j_1 m'_1, j_2 m'_2 \rangle  = -\frac{C_3(m_1m_2, m'_1m'_2)}{R^3}}
    \nonumber \\ 
         &&\quad\quad\quad\quad\quad\quad  =  \langle j_1m_1, j_2m_2 | \hat{V}_{\mu\mu}| j_1m'_1, j_2m'_2 \rangle , \label{dipole}         
\end{eqnarray}
where $\hat V_{\mu\mu}$ is magnetic dipole-dipole operator 
\begin{equation}
     \hat V_{\mu\mu} = \frac{\mu_0 (g_j \mu_B)^2}{4\pi}  \frac{(\vec j_1\cdot \vec j_2) - 3 j_{1z} j_{2z}}{R^3}\,,
\end{equation}
and $g_j=1.24159$ is the g-factor for the ground $^5{\rm I}_8$ 
state of the Dy atom \cite{NIST_g},  and $\mu_0$ is the magnetic constant. A more accurate value for the magnetic 
moment of Dy is $\mu$ = $g_j\mu_B \times j$ = 9.93 $\mu_B$.

\begin{figure} \includegraphics[scale=0.3]{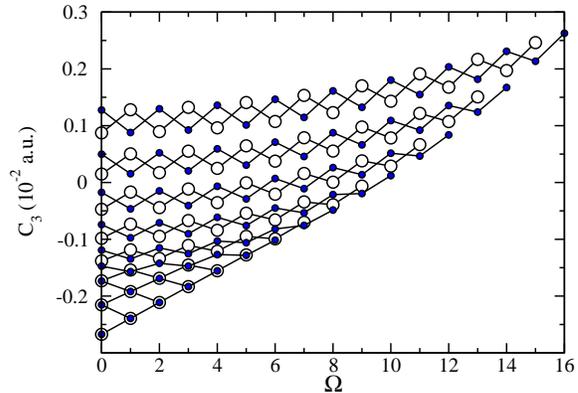}
\caption{Gerade (filled circles) and ungerade (open circles) adiabatic $C_3$ coefficients for
the interaction between two ground $^5{\rm I}_{j=8}$ state Dy atoms as a function of
the projection $\Omega$ of the total angular momentum $\vec J$ on the interatomic axis.
A larger C$_3$ coefficient implies a deeper Born-Oppenheimer potential.}
\label{C3}
\end{figure}

Figure~\ref{C3} shows the adiabatic gerade and ungerade $C_3$ coefficients as a function of $\Omega$.
These coefficients are obtained by diagonalizing the matrix Eq.(\ref{dipole}) at each $R$. The values
are both positive and negative. A comparison with the adiabatic $C_6$ coefficients in Fig.~\ref{C6} shows 
a different $\Omega$ dependence. 

\begin{figure}
\includegraphics[scale=0.3]{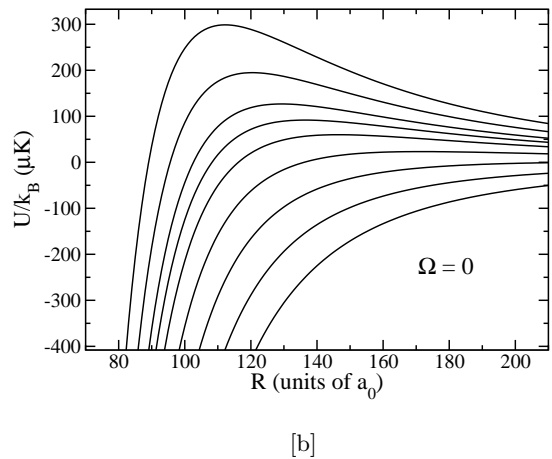} [b]
\caption{Adiabatic gerade interaction potentials of the combined electrostatic dispersion and
magnetic dipole-dipole  forces between two Dy atoms in the ground $^5{\rm I}_8$ state and projection $\Omega$=0.
Here $k_B$ is the Boltzmann constant. The effect of rotation is not included.}
\label{combined}
\end{figure}

A more accurate description of the long-range interaction is obtained by first adding the
dispersion $U_{\rm disp}$ and magnetic dipole-dipole $U_{\rm mdd}$ interactions together
and diagonalize at each internuclear separation $R$. Unlike, for the previous cases the
eigenfunctions now depend on $R$.
As an example, the resulting adiabatic gerade potentials for projection $\Omega$=0
as a function of $R$ are shown in Fig.~\ref{combined}.  At small $R$ the dispersion
interaction dominates, whereas for $R> 150 \,a_0$ the magnetic
dipole-dipole interaction plays a major role. For intermediate 
$R$ these forces compete leading to both attractive and repulsive potentials
depending on sign of the $C_3$ coefficient.  

Our unrestricted coupled cluster calculation with single, double, and perturbative triple 
excitations UCCSD(T) \cite{uccsdt} shows that  the
quadrupole moment of the Dy atom in the $^5{\rm I}_8$ state is very
small and equal to $Q$=-0.00524 a.u.. As a result the quadrupole-quadrupole
interaction energy is  seven orders of magnitude weaker than
the other atom-atom interactions. 

\section{Interactions in a magnetic trap}\label{strength}

We now analyze the relative strength of all interactions between two Dy atoms in
a magnetic field.  The magnetic field is added as either the atoms
are held in a magnetic trap \cite{BLev2010} with a spatially varying
field strength or are held in an optical trap with an homogeneous
$B$ field to control the interaction between the atoms. In addition,
the molecule can rotate, which is described by the Hamiltonian
$\hbar^2\vec \ell^2/(2m_r R^2)$, where  $\vec \ell$ is the relative orbital angular momentum between the
two atoms and $m_r$ is the reduced mass.

 It is convenient to choose a coordinate system with projection quantum numbers defined  along the external
magnetic field direction. Again following convention, projection quantum numbers are labeled by roman symbols.  
In this coordinate system the rotational and
Zeeman interactions as well as the isotropic or ``average'' dispersion potential shift molecular levels, whereas the magnetic
dipole-dipole interaction and anisotropic component of the dispersion potential lead to coupling
between different rotational and Zeeman components. As a result, the
angular momentum projection $M$ of $\vec{J}$ can change up to 2 units
due to the magnetic dipole interaction and up to 4 units due to the
anisotropic dispersion potential \cite{CGreen}.

\begin{figure}
\includegraphics[scale=0.3]{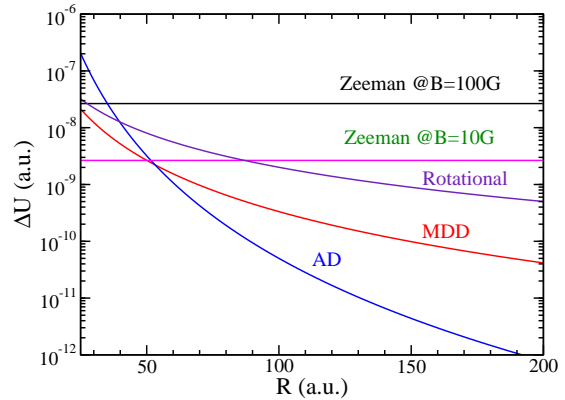}
\caption{Level splitting due to the dominant interaction forces in atomic units 
as a function of interatomic 
separation. In atomic units the Zeeman splitting is $g_jB/2$, the splitting between $\ell$=0 and 2 rotational
levels is given by $6/(2m_r R^2)$, the splitting due to the magnetic dipole-dipole 
(MDD) interaction is $2\alpha^2j(g_j/2)^2/R^3$, where $\alpha$ is the fine
structure constant, and the anisotropic dispersion (AD) 
interaction is $\Delta C_6/R^6$, where $\Delta C_6$=25 a.u.. Here $m_r$ is the reduced mass in units of the electron mass.}
\label{All}
\end{figure}

Figure~\ref{All} shows various anisotropic properties that
can lead to reorientation of the Dy angular momenta as a function
of $R$.  Firstly, the Zeeman splitting $g_j \mu_BB$ between neighboring
magnetic sublevels for magnetic field strength of 10 and 100 Gauss are
shown.  The anisotropic potential $\Delta C_6$/$R^6$ is drawn assuming
a typical value of $\Delta C_6$ = 25 a.u., based on the spread of $C_6$
value shown in Fig.~\ref{C6}.  We also present the splitting between the
rotational levels $\ell$=0 and 2 of the ground state as $6\hbar^2/(2m_r
R^2)$. Finally, the value of splitting due to magnetic dipole-dipole interaction
is given.  

For different interatomic separations different forces dominate.  In fact, when 
the curves for the magnetic dipole or anisotropic dispersion interaction
cross the Zeeman or rotational energies spin flips can occur.
At large $R$ the Zeeman splitting is dominates. Both
magnetic dipole-dipole and anisotropic electrostatic curves cross the
Zeeman $B$ =100 G curve at $R <  35a_0$, where chemical bonding should play an
important role as well.  For the weaker magnetic field of $B$ = 10 G the spin coupling occurs
for $R$ near 50$a_0$. 
The interactions will lead to mixing of rotational levels for $R<50a_0$ as well.

\begin{figure}
\includegraphics[scale=0.3]{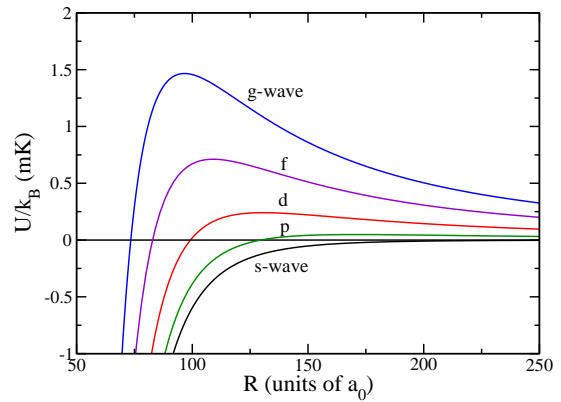}
\caption{Dispersion interaction potentials for $C_6$ = 1890 a.u. for the lowest five partial waves with centrifugal
barriers for $p$-, $d$-, $f$-, and $g$-waves}
\label{barrier}
\end{figure}

\section{First estimate of inelastic rate coefficients}\label{rates}

In this paper we complete our analysis of the interaction between
ultracold $^{164}$Dy atoms by a first estimate of inelastic loss rates
due to the anisotropy in the dispersion and magnetic dipole-dipole
interaction. We will perform a separate estimate of losses from these
interactions.  For both cases we consider atoms in a magnetic field with a strength of the
order of 1 G as described in the recent experiment \cite{BLev2010}, with
the goal to model its losses. The experiment started from a gas of Dy
atoms in a quadrupole magnetic trap with atoms distributed over atomic magnetic sublevels  with positive magnetic moment. Ie.
states $| j m\rangle$ with $m>0$ where  the projection $m$ is defined along the
magnetic field direction.  Inelastic spin-exchange collisions to states with $m\le 0$ lead
to atom loss.
 
We first describe a model for the loss due to the anisotropic dispersion potentials based on an universal single-channel scattering model developed
and used in Refs.~\cite{Mies,Julienne,Hudson}. This universal loss model assumes scattering from a single potential of the form 
$-C_6/R^6 + \hbar^2\ell(\ell+1)/(2m_r R^2)$  for $R>R_c$ and that all flux that reaches the critical separation $R_c$  undergoes irreversible 
spin-exchange independent of collision energy and partial wave $\vec \ell$. 

 For Dy we can use this universal model under several assumptions. We first note that
the anisotropy $\Delta C_6$ of the dispersion potential is small compared to the average or
isotropic dispersion potential. Secondly, an external magnetic field is applied that splits the different $m$ levels
and, as shown in Fig.~\ref{All}, the spin flip occurs between $R_c=35a_0$ and $50a_0$ depending on the magnetic field strength.
We can therefore apply the universal model assuming a mean isotropic $C_6$ value
and that, due to the anisotropic dispersion potential,  no flux returns from $R<R_c$.

\begin{figure}
\includegraphics[scale=0.3]{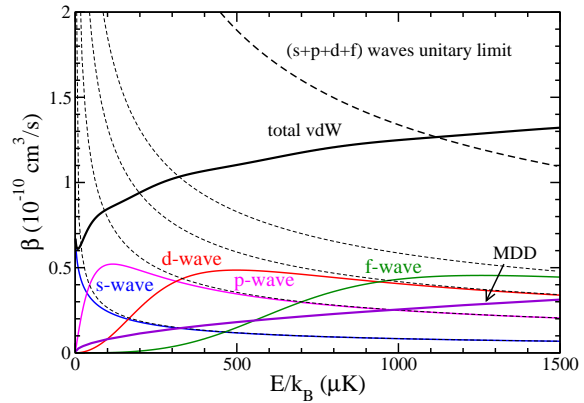}
\caption{The inelastic loss rate coefficient for a non-spin-polarized
sample of ground state $^{164}$Dy atoms as a function of collision
energy based on an universal scattering model for losses due to the anisotropy of the
dispersion potential and a Born approximation for losses from the magnetic dipole-dipole interaction. 
For the universal scattering model rate coefficients for the lowest four partial waves as well as summed rate
are shown. The loss rate coefficient for the magnetic dipole-dipole interaction is indicated by the abbreviation MDD.
The unitary limited loss rate coefficients for the lowest four partial waves are plotted as dashed lines.}
\label{InLoss}
\end{figure}

For temperatures between 100 $\mu$K to 1 mK only a few partial waves $\ell$ contribute to the
collisions. Figure~\ref{barrier} illustrates this  by showing
 the centrifugal barriers for $s, p, d, f$ and $g$ partial
waves as a function of $R$. The temperature range of interest lies
well below the $g$-wave barrier. 
Within the universal scattering model the contribution to the inelastic
rate coefficient  for partial wave $\ell$ and projection $m_{\ell}$ is 
\begin{equation}
   K_{\ell m_{\ell}}(E)= v_i\frac{\pi}{k_i^2} \left( 1- |S_{\ell m_{\ell}}(E)|^2\right) \,, 
\label{lossrate}
\end{equation}
where $E=k_i^2/(2m_r)$ is the collision energy, $k_i$ is the initial relative wavenumber, $v_i$ is the initial relative velocity, 
and the $S_{\ell m_{\ell}}(E)$ are diagonal scattering $S$-matrix elements.
The solution $\Psi_{\ell m_\ell} (R)$ of the radial Schr\"odinger equation for the single-channel potential
with boundary condition 
\[
   \Psi_{\ell m_\ell}(R) \propto e^{-i(R_x/R)^2/2} \,
\]
at short range $R<R_c$ with $R_x=\sqrt[4]{2m_rC_6/\hbar^2}$
and 
\[
   \Psi_{\ell m_\ell}(R) = \frac{e^{-ik_iR}}{\sqrt{k_i}} - S_{\ell m_{\ell}}(E)\frac{e^{ik_iR}}{\sqrt{k_i}}\,
\]
at large $R$ determines the $S_{\ell m_{\ell}}$ matrix elements. The partial and total loss
rate coefficient are $\beta_\ell(E) = 2\sum_{m_{\ell}}K_{\ell m_{\ell}}(E)$ and  $\beta(E) = 2\sum_{\ell m_{\ell}}K_{\ell m_{\ell}}(E)$, respectively. The factor of 2 is due
to the fact that after each collision two atoms are lost. 
Partial and total inelastic loss rate coefficients  are shown in Fig.~\ref{InLoss} for collision energies upto
1.5 mK. The figure shows that the loss rate for the different partial waves becomes large for collision
energies approaching the corresponding centrifugal barrier. Moreover, except for extremely small collision
energies the total loss rate coefficient slowly increases with energy.
For comparison we have also indicated the unitarity limit for each partial wave.
The unitary limit occurs when $|S_{\ell m_{\ell}}(E)|^2=0$ for all collision energies.

We now turn to a model for losses due to spin exchange induced by the magnetic
dipole-dipole interaction. For simplicity we assume that the atoms are in the stretched state
with $m=+j$.
We estimate this inelastic loss rates using first-order perturbation theory similar to that applied for
the calculation of the dipolar relaxation rates in a gas of chromium atoms
\cite{Hensler2003}. We immediately note that the magnetic moment of dysprosium or chromium
atoms is large and, therefore, a perturbative theory may not provide
an accurate loss rates. However, it is expected to give a reasonable
estimate. Following Ref.~\cite{Hensler2003}, the loss
rate coefficient for a single spin flip with $M\to M-1$ (i.e. from M=16 to 15), averaged over
all possible relative orientations of the initial relative momentum
$\vec k_i$, is
\begin{equation} \gamma_1 = \frac{4\pi}{15}j^3
\left( \frac{\mu_0  (g_j \mu_B)^2\mu_r} {2\pi \hbar^2} \right
)^2 [1+h(k_f/k_i)] \frac{\hbar k_f}{\mu_r}, 
\end{equation} 
where $\hbar^2k_f^2/(2\mu_r)=g_j\mu_B B$
and 
\[h(x)=-1/2-(3/4) (1-x^2)^2 \log[(x-1)/(x+1)]/(x(1+x^2)) 
\] 
for $x>1$.

Similarly, for a double spin
flip with $M\to M-2$ the rate coefficient is 
\begin{equation} \gamma_2 = \frac{2\pi}{15}j^2  \left( \frac{\mu_0  (g_j \mu_B)^2\mu_r} {2\pi \hbar^2}
\right )^2 [1+h(k_f/k_i)] \frac{\hbar k_f}{\mu_r}, 
\end{equation} 
where now $\hbar^2k_f^2/(2\mu_r)=2g_j\mu_B B$.  The total dipole-dipole loss
rate is given by $\gamma=2(\gamma_1+\gamma_2)$ and shown as
a function of collision energy in Fig.~\ref{InLoss}.

The loss rates in Fig.~\ref{InLoss} are smaller than the rate measured in Ref. \cite{BLev2010}
as $2.1(2)\times 10^{-10}$ cm$^3$/s for temperatures around 500 $\mu$K.
This suggests the presence of a resonance in the scattering process. In fact, flux of atoms can
return from small $R$, interfere with the incoming flux to lead to an increasing loss. We
obtained a similar effect in our analysis of the reactive collisions between two KRb 
molecules \cite{Kotochigova2010}.

\section{Conclusion}
In conclusion, we have studied the origin of the anisotropy in the
long-range interaction between ground state dysprosium atoms. This is a first
step towards a  complete multi-channel description of inelastic and elastic
collision between such atoms.
We find van der Waals coefficients by using known atomic dipole moments
and energy levels.  Our coefficients form a lower bound. We show that
the splitting between or anisotropy of the Born-Oppenheimer potentials is almost two order of
magnitude smaller than their average or isotropic potential.  In addition,
we have presented two approximate single-channel calculations to estimate inelastic losses when
Dy atoms are not in the energetically-lowest Zeeman sublevel. The first model describes losses due to
the anisotropy of the dispersion potentials and is based on an universal scattering theory. The second
perturbative model describes losses due to the magnetic dipole-dipole interaction.
The only way to obtain a clear and quantitative understanding of collisions between Dy atoms
is by a coupled-channel calculation.  We will do so in the near future. It will enable us to predict location of magnetic Feshbach
resonances in the energetically-lowest Zeeman level.

\section{Acknowledgments} This work is supported by grants of the Air
Force Office of Scientific Research and NSF PHY-1005453. We acknowledge
helpful discussions on the Dy electronic  structure with Dr. J. Reader
and stimulating discussions with Dr. B. Lev, Dr. E. Tiesinga, and Dr. J. Bohn.


\begin{references}
\bibitem{Hensler2003}S. Hensler, J. Werner, A. Griesmaier, P. O. Schmidt,
A. G\"{o}rlitz, T. Pfau, S. Giovanazzi, and K. Rzazewski, Appl. Phys. B
{\bf 77}, 765 (2003).
\bibitem{Hancox2004}C. I. Hancox, S. C. Doret, M. T. Hummon, L. Luo, and J. Doyle,
Nature {\bf 431}, 281 (2004).
\bibitem{Hancox2005}C. I. Hancox, S. C. Doret, M. T. Hummon, R. V. Krems, and J. M. Doyle,
Phys. Rev. Lett. {\bf 94}, 013201 (2005).
\bibitem{Krems2005}R.V. Krems, J. Klos, M. F. Rode, M. M. Szczesniak, G. Chalasi\'nski, 
and A. Dalgarno, Phys. Rev. Lett. {\bf 94}, 013202 (2005).
\bibitem{Stuhler2005}J. Stuhler, A. Griesmaier, T. Koch, T. Pfau, S. Giovanazzi, P. Pedri,
and L. Santos, Phys. Rev. lett. {\bf 95}, 150406 (2005).
\bibitem{Harris2007}J. G. E. Harris, S.V. Nguyen, S. C. Doret, W. Ketterle, and J. M. Doyle,
Phys. Rev. Lett. {\bf 99}, 223201 (2007).
\bibitem{Lahaye2007}T. Lahaye, T. Koch, B. Fr\"{o}hlich, M. Fattori, J. Menz, A. Griesmaier,
S. Giovanazzi, and T. Pfau, Nature {\bf 448}, 672 (2007).
\bibitem{Connolly2010}C. B. Connolly, Y. Shan Au, S. C. Doret, W. Ketterle, and J. M. Doyle,
Phys. Rev. A {\bf 81}, 01000702(R) (2010).
\bibitem{Fregoso2009}B. Fregoso and E. Fradkin, Phys. Rev. Lett. {\bf 103}, 205301 (2009);
B. Fregoso and E. Fradkin, Phys. Rev. B {\bf 81}, 214443 (2010).
\bibitem{Wilson2009}R. Wilson, S. Ronen, and J. Bohn, Phys. Rev. A {\bf 80}, 023614 (2009).
\bibitem{Baranov2011}M. A. Baranov, A. Micheli, S. Ronen, and P. Zoller, Phys. Rev. A 
{\bf 83}, 043602 (2011).
\bibitem{Wang2006}Daw-Wei Wang, Mikhail D. Lukin, and Eugene Demler, Phys. Rev. Lett.
{\bf 97}, 180413 (2006).
\bibitem{Lewenstein2006}M. Lewenstein, Nature Phys. {\bf 2} 2 (2006).
\bibitem{Aleksandrov1983}E. B. Aleksandrov {\it et al.}, Opt. Spectrosc. {\bf 54}, 1 (1983).
\bibitem{BLev2010}M. Lu, S. Ho Youn, and B. Lev, Phys. Rev. Lett. {\bf 104}, 063001 (2010).
\bibitem{Stone}A. J. Stone, {\it The theory of intermolecular forces}, (Clarendon Press, London, 1996).
\bibitem{Exp_f}M. E. Wickliffe, J. E. Lawler, and G. Nave,
J. Quantitative Spectroscopy \& Radiative Transfer {\bf 66}, 363-404 (2000); J. J. Curry, 
E. A. Den Hartog, and J. E. Lawler, J. Opt. Soc. Am. B 14, 2788 (1997);  V. N. Gorshkov, 
V. A. Komarovskii, A. L. Osherovich, N. P. Penkin, and R. Khefferlin, Opt. Spectrosc. 48, 362 (1980).
\bibitem{NIST_g}Ralchenko, Yu., Kramida, A.E., Reader, J., and NIST ASD Team (2010).
NIST Atomic Spectra Database (ver. 4.0.1), http://physics.nist.gov/asd 
\bibitem{uccsdt}J. D. Watts, J. Gauss. and R. J. Bartlett, J. Chem. Phys. {\bf 98}, 8718 (1993).
\bibitem{CGreen}R. Santra and C. H. Greene, Phys. Rev. A {\bf 67}, 062713 (2003).
\bibitem{Mies}P. S. Julienne and F. H. Mies, J. Opt. Soc. Am. B {\bf 6}, 2257 (1989).  
\bibitem{Julienne}C. Orzel, M. Walhout, U. Sterr, P. S. Julienne, and S. L. Rolston, Phys. Rev. A
{\bf 59}, 1926 (1999).
\bibitem{Hudson}E. R. Hudson, N. B. Gilfoy, S. Kotochigova, J. M. Sage, and D. DeMille,
Phys. Rev. Letter {\bf 100}, 203201 (2008).
\bibitem{Kotochigova2010}S. Kotochigova, New J. Phys. {\bf 12}, 073041 (2010). 
\end{references}
\end{document}